\documentclass[12pt]{article}
\usepackage[latin1]{inputenc}
\usepackage{graphicx}
\usepackage{float}

\begin{document}
\title{Derivation of an Abelian effective model for instanton chains in $3D$ Yang-Mills theory}
\author{~A.~L.~L.~de Lemos, ~L.~E.~Oxman, B.~F.~I.~Teixeira\\ \\Instituto de F\'{\i}sica, Universidade Federal Fluminense,\\
Campus da Praia Vermelha, Niter\'oi, 24210-340, RJ, Brazil.}
\date{\today}
\maketitle





\begin{abstract}

In this work, we derive a recently proposed Abelian model to describe the interaction of correlated monopoles, center vortices,
and dual fields in three dimensional $SU(2)$ Yang-Mills theory. Following recent polymer techniques, 
special care is taken to obtain the end-to-end probability for a single interacting center vortex,
which constitutes a key ingredient to represent the ensemble integration.

\end{abstract}

\section{Introduction}

Correlated monopoles and center vortices are believed to play a relevant role in accommodating the different properties 
of the confining string in Yang-Mills theories, receiving support from lattice simulations \cite{AGG}-\cite{GKPSZ}. 
In fact, scenarios based on either monopoles or closed center vortices are only partially successful to describe the expected 
behavior of the potential between quarks (for a review, see ref. 
\cite{greensite}). At asymptotic distances, this potential should be linear and depend on the representation of the
subgroup $Z(N)$ of $SU(N)$ ($N$-ality). At intermidiate scales, it should posses Casimir scaling.

Monopoles can be seen as defects that arise when implementing the Abelian projection \cite{ap}. 
The Cho-Faddeev-Niemi representation (CFN) can also be used to associate monopoles with defects of the local 
color frame used to decompose the gauge fields  \cite{cho-a}-\cite{Shaba}. An important issue is how to deal with nonphysical 
objects such as Dirac strings (or worldsheets) when charged fields are present. This has been studied in ref. \cite{andre},
using the CFN representation of $SU(2)$ Yang-Mills theory. There, we showed how to decouple Dirac strings in the partition function of the theory, only
leaving the effect of their borders where monopoles are placed. 

In ref. \cite{oxman}, the possible frame defects were extended to describe 
not only monopoles but also center vortices, correlated or not. In ref. \cite{oxman2}, this procedure has been related 
with the usual manner to introduce thin center vortices in the continuum, providing a natural manner to
discuss the stability of center vortices. In this framework, we also discussed the relationship between large dual transformations in three and 
four dimensional Yang-Mills theories and phases where Wilson surfaces can be either decoupled or become integration variables \cite{ldual}. This is relevant 
for the problem of confinement, a phase where the surface whose border is the Wilson loop becomes observable. 

In these scenarios, one of the difficulties is how to deal 
with the integration over an ensemble of extended objects, after considering a phenomenological parametrization of their properties, such as stiffness, 
interactions with dual fields, and interactions between them. This is particularly severe in four dimensional theories where center vortices generate two 
dimensional extended worldsurfaces. However, in three dimensions center vortices are stringlike and an ensemble of 
worldlines is naturally associated with a second quantized field theory. For this reason, we were able to propose in refs. \cite{oxman,ldual}, 
following heuristic arguments, an Abelian effective model to describe the large distance behavior of 
the 3D $SU(2)$ Yang-Mills theory (for a non Abelian version, see ref. \cite{CL}). This model corresponds to a generalization of the t' Hooft model \cite{hooft1};
it includes a coupling with the dual field that can be defined in order to represent the off-diagonal sector. This coupling 
is essential to relate the possible vortex phases with enabled or disabled large dual transformations, and discuss in this framework 
the observability of Wilson surfaces \cite{ldual}. 

The aim of this article is presenting a careful derivation of this effective model, after parameterizing some 
intrinsic physical properties that these objects could present. 
One of the fundamental ingredients will be the adoption of recent techniques borrowed from polymer physics \cite{fred}, 
where the extended objects are also one dimensional. The polymer formulation of field theory in Euclidean spacetime \cite{Ambjorn} 
has also been used to study the magnetic component of the Yang-Mills plasma due to monopoles \cite{CZ}, 
which in four dimensional spacetime are stringlike objects. 

In this work, we present a detailed derivation of the equation for the end-to-end probability for a center vortex worldline, 
including the effect of interactions. This probability can be thought of as a Green's function that depends on the position and orientation at the 
worldline boundaries, where monopole-like instantons are placed. In the limit of semiflexible polymers, a reduced Green's function for a complex vortex field minimally 
coupled to the dual field is obtained. This constitutes a key ingredient to derive the above mentioned effective model.

In section \S \ref{review}, we briefly review how to use the CFN decomposition to include vortices as defects of the local color frame.
In \S \ref{Qcorr}, we rewrite the ensemble for correlated monopoles and center vortices in terms of the weight for a single interacting vortex,
while in \S \ref{weight} we derive the associated Fokker-Planck equation. In section \S \ref{emodel}, we combine the previous results to obtain the 
generalized effective model. Finally, in \S \ref{conc} we present our conclusions.

\section{Correlated instantons and center vortices}
\label{review}

The starting point is the $SU(2)$ Yang-Mills action defined in three dimensional
Euclidean spacetime,
\begin{equation}
{S_{YM}}=\frac{1}{2}\int d^3 x\; tr\, (F_{\mu \nu} F_{\mu \nu })
\makebox[.5in]{,}
F_{\mu \nu}=F_{\mu \nu}^{a}T^{a}.
\label{a1}
\end{equation}
The generators of $SU(2)$ can be expressed in terms of Pauli matrices
$T^a=\tau^a/2$, $a=1,2,3$, and the field strength in terms of the gauge fields
$A_{\mu }^{a}$,
\begin{equation}
\vec{F}_{\mu \nu}=\partial_\mu \vec{A}_\nu -\partial_\nu \vec{A}_\mu +g
\vec{A}_\mu\times \vec{A}_\nu,
\makebox[.5in]{,}
\vec{A}_\mu=A_\mu^a\, \hat{e}_a
\makebox[.5in]{,}
\vec{F}_{\mu \nu}=F_{\mu \nu}^a\, \hat{e}_a,
\label{a2}
\end{equation}
where $\hat{e}_a$ is the canonical basis in color space, $a=1, 2, 3$.

Following ref. \cite{oxman}, in order to separate the perturbative sector from the sector of topological 
defects, we can use the Cho-Fadeev-Niemi representation,
\begin{equation}
\vec{A}_\mu = \biggl[A_{\mu}\hat{n} - \frac{1}{g}
\hat{n}\times\partial_{\mu} \hat{n}+ X_{\mu}^{1}\hat{n}_{1} +
X_{\mu}^{2}\hat{n}_{2}\biggr].
\label{a5}
\end{equation}
Here, the fields $X^1_\mu$, $X^2_\mu$ represent off-diagonal fluctuations, while correlated
monopoles and center vortices can be parametrized as defects of the local color frame $\hat{n}_{a}$ 
($\hat{n}_3 \equiv \hat{n}$). In this scenario, the obtained Yang-Mills partition function is \cite{oxman},
\begin{equation}
Z_{YM} = \int [{\cal D}\lambda][\mathcal{D}\Psi]\, e^{-S_{c}- \int d^{3}x\,
\frac{1}{2}\lambda_{\mu}\lambda_{\mu}} e^{ i\int d^{3}x [\lambda_{\mu}k_{\mu}
+A_{\mu}
(\epsilon_{\mu\nu\rho}\partial_{\nu}\lambda_{\rho} - J_{\mu}^{c})]}\, Z_{v,m},
\label{W3d}
\end{equation}
where $ k_\mu=\frac{g}{2i}\epsilon_{\mu \nu \rho}(\bar{\Phi}_\nu \Phi_\rho- \Phi_\nu
\bar{\Phi}_\rho)$, and $S_{c}$ is the action for the charged fields,
\begin{equation}
S_c=\int d^3x\, \left[\bar{\Lambda}^{\mu} \Lambda^{\mu}-i (\bar{\Lambda}^{\mu } 
\epsilon^{\mu \nu \rho }\partial_\nu \Phi_\rho + {\Lambda}^{\mu } 
\epsilon^{\mu \nu \rho }\partial_\nu \bar{\Phi}_\rho)\right],
\end{equation}
$\Phi_\mu =(X_\mu^1+iX_\mu^2)/\sqrt{2} $. 
The measure $[\mathcal{D}\Psi]$ integrates over gluon, ghost and auxiliary fields, and
the conserved color current can be written in the form $J_{\mu}^{c} = J^\mu +K^\mu$, with 
$J^\mu= ig \epsilon^{\mu \nu \rho} \bar{\Lambda}_{\nu }\Phi_\rho - ig\epsilon^{\mu \nu \rho } {\Lambda}_{\nu }\bar{\Phi}_\rho$, and  
$K^\mu$ receiving the contribution from charged fields of the gauge fixing sector. Note that in eq. (\ref{W3d}) we have the implicit constraint,
\begin{equation}
J^c_\mu=\epsilon_{\mu \nu \rho} \partial_\nu \lambda_{\rho},
\label{const-3d}
\end{equation}
that is, the topologically conserved current associated with $\lambda_\mu$ describes the off-diagonal sector.
The partition function for the correlated monopoles and center vortices can be written in the form, 
\begin{equation}
Z_{v,m} =\int [Dm][Dv]\, e^{i\int d^{3}x \lambda_{\mu}
d_{\mu}}.
\label{a14}
\end{equation}
The measure $[Dm][Dv]$, representing the integration over the ensemble 
of mon\-o\-pole chains, will be specified in the following section. With regard to 
$d_{\mu}$, it is concentrated on the defects and is obtained from the defining equations,
\begin{equation}
h_{\mu}=\tilde{h}_{\mu}+ d_{\mu},
\label{mv-sep-3}
\end{equation}
\begin{equation}
h_{\mu}=-\frac{1}{2g}\epsilon_{\mu \nu \rho}\,
\hat{n}\cdot(\partial_\nu\hat{n}\times \partial_\rho\hat{n})
\makebox[.3in]{,}
\tilde{h}_{\mu}=\epsilon_{\mu \nu \rho}\partial_\mu C_\rho
\makebox[.3in]{,}
C_\mu=-\frac{1}{g} \hat{n}_1 \cdot \partial_\mu \hat{n}_2. 
\end{equation}

As an example, for a monopole/anti-monopole pair correlated with center vortices, we have,
\begin{equation}
d_{\mu}=d_{\mu}^{(1)}+d_{\mu}^{(2)}
\makebox[.5in]{,}
d_\mu^{(\alpha)} =\frac{2\pi}{g} \int d\sigma\,
\frac{dx^{\alpha}_\mu}{d\sigma}\, \delta^{(3)}(x-x^\alpha(\sigma)).
\end{equation}
Here, $x^\alpha(\sigma)$, $\alpha=1,2$, is
a pair of open center vortex
worldlines with the same boundaries at $x^+$, $x^-$, where the monopole and anti-monopole are localized, 
so that it is verified,
\begin{equation}
\partial_\mu d_\mu^{(\alpha)}
=\frac{2\pi}{g}(\delta^{(3)}(x-x^+)-\delta^{(3)}(x-x^-)).
\label{divg}
\end{equation}

\section{Ensemble of instanton chains}
\label{Qcorr}

To start handling the ensemble integration over defects, we write the partition function for the monopole chains in the form,
\begin{eqnarray}
\lefteqn{Z_{v,m}=\int [\mathcal{D}\phi]\, e^{-S_\phi}\, \sum_n\, Z_n}, \nonumber \\
&& Z_n= \int [\mathcal{D}m]_n [\mathcal{D}v]_n \, \exp \left[ {\,i \frac{2\pi}{g}\sum_{k=1}^{2n}\int_{0}^{L_{k}} ds\; 
\dot{x}^{(k)}_\alpha \lambda_\alpha(x^{(k)}) - S_{d}}\right],
\label{Zn}
\end{eqnarray}
\begin{equation}
S_{d}=\sum_{k=1}^{2n}{\int_{0}^{L_{k}} ds\, \biggl[ \mu + \frac{1}{2\kappa}\, \dot{u}^{(k)}_\alpha \dot{u}^{(k)}_\alpha + 
\phi(x^{(k)})\biggr]}.
\nonumber
\label{phenom}
\end{equation}
The integer $n$ denotes the number of instanton/anti-instanton pairs. Center vortices are attached in pairs to the previous pointlike 
objects, so that for a given realization of defects, with a given $n$, the number of attached center vortex worldlines is $2n$. In the previous formula 
these stringlike objects has been denoted by $x^{(k)}(s)$, $k=1,\dots,2n$. For each center vortex, $s$ denotes the associated arc length parameter 
running from $0$ to $L_k$, the total length of the worldline. In terms of the tangent vector $u^{(k)}(s) = \dot{x}^{(k)}(s)$, the defining condition for this parameter is 
$u^{(k)}_\alpha u^{(k)}_\alpha =1$, where $\alpha$ is summed over $\alpha=1,2,3$ (no summation over $k$).

In eq. (\ref{phenom}), we have the phenomenological terms containing dimensional parameters. 
The first term in $S_{d}$ describes tensile center vortices, the second one is associated with their stiffness. Note that using the density 
$\rho(x) = \sum_{k}{\int_{0}^{L_{k}} ds\, \delta\left(x- x^{(k)}(s)\right)}$, if the  path integral over $\phi$ were performed with,
\begin{equation}
e^{-S_{\phi}} = e^{\frac{1}{2}\int d^{3}x\, d^{3}x'\, \phi(x) V^{-1}(x,x')\phi(x')},
\nonumber
\end{equation}
then the interaction factor between center vortices would be obtained,
\begin{equation}
e^{-\frac{1}{2}\int d^{3}x\, d^{3}x'\, \rho(x) V(x,x')\rho(x')}
\makebox[.3in]{,}
\int d^3y\, V^{-1}(x,y) V(y,z)=\delta(x-z) .
\end{equation} 
In particular, taking $ V(x-y)= (1/\zeta)\, \delta(x-y)$, in which case,
\begin{equation}
 e^{-S_{\phi}} = e^{\int d^{3}x\, \frac{\zeta}{2}\, \phi^2},
\end{equation}
corresponds to a contact interaction. For a given $n$, the measure $[\mathcal{D}m]_n =\xi^{2n}\,  d^3 x_1 \dots d^3 x_{2n}$ represents the integral over the positions of 
the $2n$ instantons and anti-instantons. The parameter $\xi$ has dimension of mass, and is necessary to match the dimensions of the different terms.
For a given realization of the monopole positions, the $[\mathcal{D}v]_n$ integration measure includes the 
sum on the different inequivalent manners to join them with center vortices, with the associated symmetry factor. In addition,
for each one of the $2n$ center vortices, this measure contains the path integral over all center vortex worldlines 
$[Dx^{(k)}(s)]$ with fixed extrema, and fixed length $L_k$, followed by the integral over the lengths $ \int_0^{\infty} dL_k$.

Then, from eq. (\ref{Zn}), it becomes clear that all possible terms in the partition function depend on a fundamental 
building block, namely, the weight associated with center vortices with fixed endpoints, 
\begin{eqnarray}
\lefteqn{Q(x,x_0)= \int_0^{\infty} dL\, e^{-\mu L}\, q(x,x_0,L),}\nonumber \\
&& q(x,x_0,L)= \int [Dx(s)]\, e^{ -\int_{0}^{L} ds\, \left[\frac{1}{2\kappa}\, \dot{u}_\alpha \dot{u}_\alpha
+\phi(x(s))-i \frac{2\pi}{g} u_\alpha(s) \lambda_\alpha(x(s))\right]},
\label{Qprin}
\end{eqnarray}
where $[Dx(s)]$ represents the integral over all possible paths $x(s)$ with fixed length $L$, and extrema at $x_0$ and $x$.

\begin{figure}[H]
\centering
\includegraphics[scale=0.15, bb = 0 0 600 350]{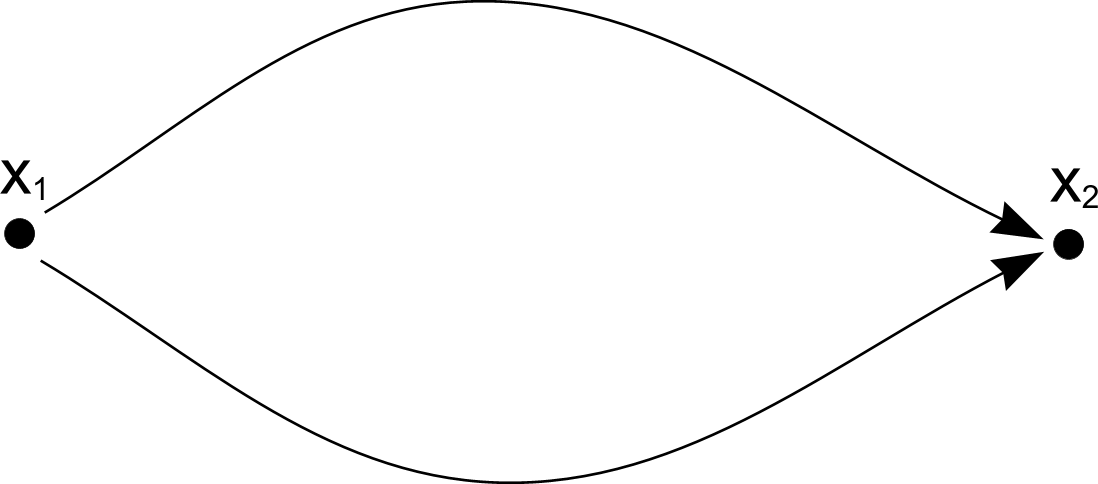}
\caption{Instanton/anti-instanton correlated with a pair of center vortices}
\label{one}
\end{figure}

\begin{figure}[H]
\centering
\includegraphics[scale=0.4, bb = 0 0 800 350]{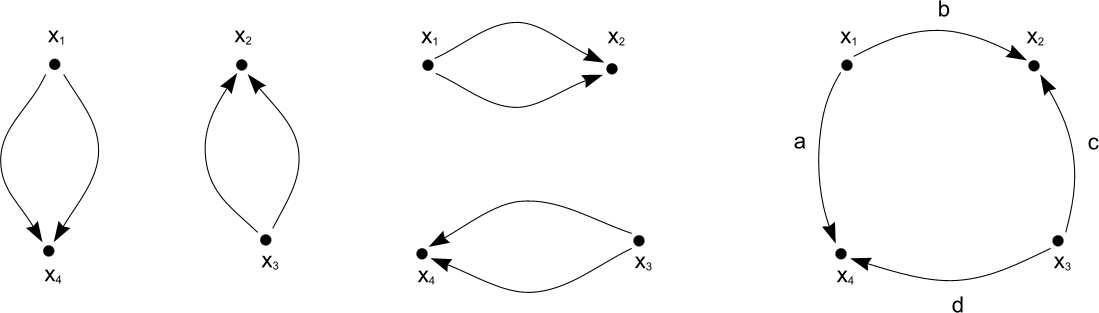}
\caption{Different manners to correlate two given instanton/anti-instanton pairs with center vortices.}
\label{two}
\end{figure}

For an instanton/anti-instanton pair (fig. \ref{one}), we have the contribution:
\begin{equation}
Z_{1} = \frac{1}{2!} \int d^3 x_1 d^3 x_2\; \xi^2 \left[ Q^{2}_{x_{2},x_{1}} + 
Q^{2}_{x_{1},x_{2}} \right].
\nonumber
\end{equation}
For two instanton/anti-instanton pairs, we have six different manners to
distribute instantons and anti-instantons at positions $x_{1}$, $x_{2}$, $x_{3}$
e $x_{4}$. These fixed boundaries can be linked in three different forms: 
two disconnected and one connected (fig. \ref{two}).
Note that for the connected configurations we have to consider some symmetry
aspects. We can generate a new contribution by interchanging the vortices
$a$, $b$, as well as the vortices $c$, $d$, that is, we have $2!.2!=4$ manners to realize a given connected
configuration. Then, for two pairs the contribution is,
\begin{eqnarray}
Z_{2} &=& \frac{1}{4!}\int d^3 x_1 d^3 x_2 d^3 x_3 d^3 x_4\; \xi^4 \left[ Q^{2}_{x_{4},x_{1}} Q^{2}_{x_{2},x_{3}} + 
Q^{2}_{x_{2},x_{1}} Q^{2}_{x_{4},x_{3}}  \right.
\nonumber \\
&& \left. + 4\, Q_{x_{2},x_{1}} Q_{x_{4},x_{1}} Q_{x_{2},x_{3}} Q_{x_{4},x_{3}}+ {\rm permutations} \right].
\nonumber \\
\label{2pares}
\nonumber
\end{eqnarray}
We can continue analyzing the different terms in the expansion, to obtain that all the terms can be
obtained from a functional generator as follows,
\begin{eqnarray}
\lefteqn{1+Z_{1}+Z_2+\dots = } \nonumber \\
&& \left\{ 1+ \int d^3x_1\, I\left(\frac{\delta~~}{\delta
C(x_1)}\right)+\frac{1}{2!} \int d^3x_1 d^3x_2\, I\left(\frac{\delta~~}{\delta
C(x_1)}\right) I\left(\frac{\delta~~}{\delta C(x_2)}\right)+ \right.  \nonumber \\
&&+\frac{1}{3!} \int d^3x_1 d^3x_2 d^3x_3\, I\left(\frac{\delta~~}{\delta
C(x_1)}\right) I\left(\frac{\delta~~}{\delta
C(x_2)}\right)I\left(\frac{\delta~~}{\delta C(x_3)}\right)+ \nonumber \\
&&+\left. \frac{1}{4!} \int d^3x_1 d^3x_2 d^3x_3 d^3x_4\,
I\left(\frac{\delta~~}{\delta C(x_1)}\right) I\left(\frac{\delta~~}{\delta
C(x_2)}\right)I\left(\frac{\delta~~}{\delta C(x_3)}\right)
I\left(\frac{\delta~~}{\delta C(x_4)}\right) +  \right.
\nonumber
\\
&&+\left ....\biggr\} \right.\left. e^{-\int d^3x d^3y\, \bar{K}(x) Q(x,y)K(y)}\right|_{C=0}.
\label{exp}
\end{eqnarray}
Here, we have defined the operator, 
\begin{equation}
I\left(\frac{\delta~~}{\delta C(x)}\right)= \xi
\left(\left[\frac{\delta~~}{\delta
\bar{K}(x)}\right]^2+\left[\frac{\delta~~}{\delta K(x)}\right]^2\right),
\end{equation}
where $C(x)$ represents the set of sources $K(x)$, $\bar{K}(x)$.
This can be verified by performing the functional derivatives and evaluating at $K(x)=0$, $\bar{K}(x)=0$. 
In other words, we can write, 
\begin{eqnarray}
Z_{v,m}&=&\int [\mathcal{D}\phi]\, e^{-S_\phi}\, e^{\int d^3x\, I\left(\frac{\delta~~}{\delta C(x)}\right)} \left. e^{-\int d^3x d^3y\, \bar{K}(x) Q(x,y)K(y)}\right|_{C=0}.
\label{czero}
\end{eqnarray}
Then, it becomes clear that in order to obtain an effective vortex theory, it is essential to have a simple field representation 
for the $Q$-dependent factor, thus enabling the possibility of performing the path integral over $\phi$.

\section{Statistical weight for a single center vortex}
\label{weight}

The discussion about how to represent the path-integration over a string-like object with stiffness is not simple
even in the noninteracting case. It is usually done relying on the assumption that stiffness is equivalent to 
an effective monomer size in the random chain calculation, as it tends to locally straighten the chain, which
is supported after cumbersome calculations of different momenta for the associated probability distributions \cite{kleinert}. 
For noninteracting random chains, the end-to-end probability is given by,
\begin{eqnarray}
q_{N}(x,x_{0})=\prod_{n=1}^{N}\Bigg[\int (d^{3}\Delta x_{n})\, \frac{1}{4\pi a^{2}}
\delta(|\Delta x_{n}|-a)\Bigg]\delta (x-x_{0}-\sum_{n=1}^{N}\Delta x_{n}),
\nonumber
\end{eqnarray}
which for large $N$ behaves like, 
\begin{equation}
q_{N}(x,x_{0})\approx\left(\frac{3}{2\pi Na^{2}}\right)^{3/2}\exp\left[-
\frac{3(x-x_{0})^{2}}{2Na^{2}}\right].
\end{equation}
Note that the continuum limit with $Na=L$ cannot be implemented here. However, by considering the above mentioned effective 
monomer size $a\to a_{\rm eff}$, and replacing $Na^2/3 \to L/\alpha$, $\alpha=3/a_{\rm eff}$, it results,
\begin{equation}
  q(x,x_0,L)= \left(\frac{\alpha}{2\pi L}\right)^{3/2} e^{-\frac{\alpha}{2L}(x 
- x_{0})^{2}}=\int \frac{d^{3}k}{(2\pi)^{3}}\, e^{-\frac{L}{2\alpha} k^2}\, e^{ik\cdot (x- x_{0})}, 
\end{equation}
and integrating over the different lengths, weighted by $e^{-\mu L}$, as is well-known, 
$Q(x,x_0)$ turns out to be the Green's function for a free field theory, 
\begin{equation}
Q(x,x_0) =2\alpha \int \frac{d^{3}k}{(2\pi)^{3}}\, 
\frac{e^{ik\cdot (x- x_{0})}}{k^{2} + m^{2}}.
\makebox[.5in]{,}
m^2=2\alpha \mu,
\end{equation}
\begin{equation}
 (-\nabla^2 + m^2)\, Q(x,x_0) = 2\alpha\, \delta (x-x_0).
\end{equation}

\begin{figure}[H]
\centering
\includegraphics[scale=0.3, bb = 0 0 650 150]{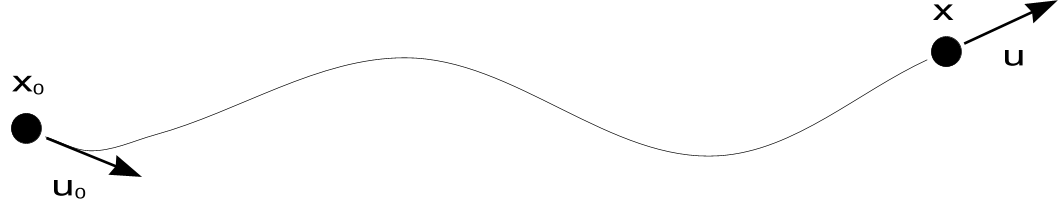}
\caption{Interacting center vortices with fixed length, and orientations at the endpoints, 
define the weight $q(x,u,x_0,u_0,L)$.}
\label{qfig}
\end{figure}

Now, we would like to present a careful extension of this property, as controlled as possible, to the case where scalar $\phi$-interactions 
and vector $\lambda$-interactions are present, as is the case of our path integral over a single center vortex in eq. (\ref{Qprin}).
In this case, the momenta of the distribution for general external sources cannot be computed in a closed form, nor an explicit expression for
the random chain integration is available. A manner to overcome this difficulty is noting that we are only interested in obtaining 
a field representation for $Q(x,x_0)$. Then, we can follow recent techniques  \cite{fred} 
for semiflexible interacting polymers, adapted to the fixed extrema and variable length stringlike objects we have in $Q(x,x_0)$. 
The desired representation, will be obtained from,
\begin{equation}
 Q(x,x_0)=\int_0^{\infty} dL\, e^{-\mu L} \int \frac{d^2u_0}{4\pi}\, \frac{d^2u}{4\pi}\, q(x,u,x_0,u_0,L),
\label{q1}
\end{equation}
where $q(x,u,x_0,u_0,L)$ is the correlator for center vortices with fixed length, positions and tangent vectors at the  edges, 
where monopoles are placed (see fig. \ref{qfig}).
The differentials $d^2u_0$, $d^2u$ integrate on the unit sphere $S^2$ and are normalized such that, $\int d^2 u_0 =\int d^2 u = 4\pi$. 

In order to generate a discretized version of $q(x,u,x_0,u_0,L)$, let us start by considering points $x \in {R}^{3}$, $u\in S^{2}$, 
the initial condition, 
\begin{equation}
q_0(x,u,x_0,u_0)=\delta(x-x_{0})\delta (u-u_{0})\, e^{-\omega(x,u)},
\label{1}
\end{equation}
where $\delta (u-u_{0})$ is defined on $S^2$, and the recursive relation,
\begin{equation}
q_{j+1}(x,u,x_0,u_0)=e^{-\omega(x,u)} \int d^3x' d^2u'\, \phi(u-u')
\delta(x - x' - u\Delta L)\, q_j(x',u',x_0,u_0).
\label{2}
\end{equation}
Using $j=0$, together with eq. (\ref{1}), we get
\begin{equation}
q_1(x,u,x_0,u_0)
= e^{-[\omega(x,u) +\omega(x_{0},u_{0})]}\phi(u-u_{0})
\delta(x - x_{0} - u\Delta L).
\label{5}
\end{equation}
Continuing the iteration, it is easy to see that for $j=N-1$, we will have,
\begin{eqnarray}
\lefteqn{q_N(x,u,x_0,u_0)=\int d^3x_{1} d^2u_{1}... d^3x_{N-1} d^2u_{N-1}\,
 }
\nonumber
\\
&& e^{-[\omega(x_{0},u_{0})+...\omega(x_{N-1},u_{N-1}) + \omega(x,u)]} \,\phi(u_{1}-u_{0}) \dots  \phi(u_{N-1}-u_{N-2})\phi(u-u_{N-1})
\times  
\nonumber
\\
&&\times \delta(x_{1} - x_{0} - u_{1}\Delta L) 
\dots  \delta(x_{N-1} - x_{N-2} - u_{N-1}\Delta L) \delta(x - x_{N-1} - u\Delta L).\nonumber \\
\label{8}
\end{eqnarray}
Defining $x=x_{N}$, $u=u_{N}$, we can rewrite eq. (\ref{8}) in a more compact form,
\begin{eqnarray}
\lefteqn{q_N(x,u,x_0,u_0)=\int d^3x_{1} d^2u_{1}\dots d^3x_{N-1} d^2u_{N-1}\, e^{-\sum_{i=0}^{N}\omega(x_{i}, u_{i})}\times}  \nonumber \\
&&\times \prod_{j=0}^{N-1} \phi(u_{j+1}-u_{j})
\delta(x_{j+1} - x_{j} - u_{j+1}\Delta L).
\label{9}
\end{eqnarray}
Then, choosing the normalized angular distribution and interaction function, 
\begin{equation}
\phi(u-u') = {\cal N} e^{-\frac{1}{2\kappa} \Delta L \left( \frac{u - u'}{\Delta L}\right)^{2}},
\label{3}
\end{equation}
\begin{equation}
\omega(x,u)=\Delta L \, \left[ \phi(x)-i\frac{2\pi}{g} u\cdot \lambda (x) \right],
\end{equation}
it becomes clear that eq. (\ref{9}) corresponds to a discretization of $q(x,u,x_0,u_0,L)$ by $N$ ``monomers", corresponding to integrate over
center vortices with the conditions $x(0)=x_{0}$, $x(L)=x$, $u(0)=u_{0}$, $u(L)=u$.
That is, 
$q(x,u,x_0,u_0,L)=\lim_{N\to \infty} q_N(x,u,x_0,u_0)$, with $L=N\Delta L$.
In addition, from eq. (\ref{2}), we have,
\begin{equation}
q_{N+1}(x,u,x_0,u_0)=e^{-\omega(x,u)} \int d^3x' d^2u'\, \phi(u-u')\,
q_N(x- u\Delta L,u',x_0,u_0),\nonumber \\
\label{10}
\end{equation}
so that for large $N$, after expanding both members in $\Delta L =L/N$, and using that $\phi(u-u')$ is localized, to expand 
$q_N(x-u\Delta L, u',x_0,u_0)$ around $u'\approx u$, the following diffusion equation is obtained (see ref. \cite{fredbook}),
\begin{eqnarray}
\partial_L \, q(x,u,x_0,u_0,L)=\left[\frac{\kappa}{2}\nabla_{u}^{2} - \phi(x) -u\cdot D_x\right] q(x,u,x_0,u_0,L).
\label{eqq}
\end{eqnarray}
Here, $\nabla^{2}_{u}$ is the Laplacian on the unit sphere, $D_x=\nabla_{x}-i\frac{2\pi}{g} \lambda (x)$, 
and the $\Delta L \to 0$ limit of eq. (\ref{1}) implies that 
this equation has to be solved with the condition,
\begin{equation}
 q(x,u,x_0,u_0,0)=\delta (x-x_0) \delta (u-u_0).
\label{initq}
\end{equation}

In the process of obtaining $Q(x,x_0)$ from $q(x,u,x_0,u_0,L)$, the integrals in eq. (\ref{q1}) can be organized as follows. We will initially
integrate over $d^2u_0$ to obtain the reduced Green's function,
\begin{equation}
q(x,u,x_0,L)=\int \frac{d^2u_0}{4\pi}\, q(x,u,x_0,u_0,L),
\end{equation}
which after integrating both members in eqs. (\ref{eqq}) and (\ref{initq}), satisfies,
\begin{eqnarray}
\partial_L \, q(x,u,x_0,L)=\left[\frac{\kappa}{2}\nabla_{u}^{2} - \phi(x) -u\cdot D_x\right] q(x,u,x_0,L),
\end{eqnarray}
\begin{equation}
q(x,u,x_0,0)=\delta (x-x_0).
\end{equation}
Next, by integrating over the different lengths, we obtain,
\begin{equation}
Q(x,u,x_0)= \int_0^{\infty} dL\, e^{-\mu L} q(x,u,x_0,L).
\end{equation}
This function verifies,
\begin{eqnarray}
\lefteqn{\left[\frac{\kappa}{2}\nabla_{u}^{2} - \phi(x) -u\cdot D_x\right]Q(x,u,x_0)=\int_0^{\infty} dL\, e^{-\mu L} 
\partial_L\, q(x,u,x_0,L)}\nonumber \\
&& =  \int_0^{\infty} dL\, \partial_L\left[e^{-\mu L} 
 q(x,u,x_0,L)\right]+ \mu\, \int_0^{\infty} dL\, e^{-\mu L}
q(x,u,x_0,L)\nonumber \\
&&=-  q(x,u,x_0,0)+ \mu\, Q(x,u,x_0),
\end{eqnarray}
that is,
\begin{equation}
\left[-\frac{\kappa}{2}\nabla_{u}^{2} + \phi(x) + u\cdot D_x +\mu\right]Q(x,u,x_0)=  \delta(x-x_0).
\end{equation}
Finally, we can obtain $ Q(x,x_0)$ from,
\begin{equation}
Q(x,x_0)= \int \frac{d^2u}{4\pi}\, Q(x,u,x_0).
\end{equation}
In other words, $Q(x,x_0)$ is given by the zeroth component ${\cal Q}_0$ of an $u$-expansion of $Q(x,u,x_0)$ in terms of different angular momenta,
\begin{equation}
Q(x,u,x_0)=\sum_{l=0} {\cal Q}_l (x,u,x_0)
\makebox[.5in]{,}
Q(x,x_0)= {\cal Q}_0(x,x_0).
\end{equation}
We can also use the expansion,
\begin{equation}
u\cdot D_x Q(x,u,x_0)= \sum_{l=0} u\cdot D_x {\cal Q}_l = \sum_{l=0} {\cal R}_l,
\end{equation}
\begin{eqnarray}
{\cal R}_0 &=& [ u\cdot D_x {\cal Q}_1 ]_0 \nonumber \\
{\cal R}_1 &=& [u\cdot D_x {\cal Q}_0  + u\cdot D_x {\cal Q}_2]_1\nonumber \\
{\cal R}_2 &=& [u\cdot D_x {\cal Q}_1 + u\cdot D_x {\cal Q}_3]_2~\dots
\label{erres}
\end{eqnarray}
to obtain,
\begin{equation}
[\phi(x)+\mu]\, {\cal Q}_0 +  {\cal R}_0 =  \delta(x-x_0),
\label{RQ0}
\end{equation}
and for $l\neq 0$,
\begin{equation}
\frac{1}{f_l(x)}\, {\cal Q}_l +  {\cal R}_l = 0 
\makebox[.5in]{,}
f_l(x)=\left[\phi(x)+\mu + \frac{ l(l+1)\kappa}{2} \right]^{-1}.
\end{equation}
Then, we have,
\begin{eqnarray}
\lefteqn{{\cal R}_0 = [u\cdot D_x {\cal Q}_1 ]_0 = -\left[u\cdot D_x (f_1 {\cal R}_1) \right]_0} \nonumber \\
&& = -\left[{\cal R}_1\, u\cdot \nabla_x f_1 + f_1\, u\cdot D_x {\cal R}_1 \right]_0 ,
\end{eqnarray}
and as in the second line of eq. (\ref{erres}), $[u\cdot D_x {\cal Q}_0]_1 = u\cdot D_x {\cal Q}_0$, we obtain,
\begin{eqnarray}
{\cal R}_0 &=&  -[(u\cdot D_x {\cal Q}_0)\, (u\cdot \nabla_x f_1) + f_1\, (u\cdot D_x)^2\,  {\cal Q}_0 ]_0 -\nonumber \\
&&[ [u\cdot D_x {\cal Q}_2]_1\, (u\cdot \nabla_x f_1 ) + f_1\, u\cdot D_x [u\cdot D_x {\cal Q}_2]_1 ]_0 .
\end{eqnarray}
Now, if the ${\cal Q}_l$ components with momentum $l\geq 2$ are supposed to be small (semiflexible limit), we get,
\begin{equation}
{\cal R}_0 \approx  - [D_\alpha {\cal Q}_0\, \partial_\beta f_1 + f_1\, D_\alpha D_\beta\,  {\cal Q}_0] [u_\alpha u_\beta]_0 ,
\end{equation}
and decomposing the tensor into a traceless symmetric ($l=2$) and scalar ($l=0$) part, 
$u_\alpha u_\beta = \left(u_{\alpha}u_{\beta}-\frac{1}{3}\delta_{\alpha\beta}\right) + \frac{1}{3}\delta_{\alpha\beta}$,
we get,
\begin{equation}
{\cal R}_0 \approx  - \frac{1}{3}[\partial_\alpha f_1 \, D_\alpha {\cal Q}_0 + f_1\, D^2\,  {\cal Q}_0],
\end{equation}
and replacing in eq. (\ref{RQ0}),
\begin{equation}
 - \frac{1}{3}[\partial_\alpha f_1 \, D_\alpha {\cal Q}_0 + f_1\, D^2\,  {\cal Q}_0] +[\phi(x)+\mu]\, {\cal Q}_0 =  \delta(x-x_0),
\end{equation}
\begin{equation}
f_1(x)=\left[\phi(x)+\mu + \kappa \right]^{-1}.
\end{equation}
Therefore, for $\kappa$ much larger than $\mu$ and the mass scales associated with $\phi$, we finally obtain the 
approximated differential equation,
\begin{equation}
 \left[- \frac{1}{ 3\kappa} D^2 + \phi(x) + \mu \right]\, {\cal Q}_0 =   \delta(x-x_0).
\end{equation}

\section{Effective Field Theory}
\label{emodel}

As a consequence of the calculations presented in the previous section, we see that the $Q$-dependent factor in the partition function $Z_{v,m}$ 
in eq. (\ref{czero})  
can be expressed in terms of a complex field $v$,
\begin{equation}
e^{-\int d^3x d^3y\, \bar{K}(x) Q(x,y)K(y)}= \det \hat{O} \int [{\cal D}v][{\cal D}\bar{v}]\, e^{-S_v-\int d^3x\, [\bar{K}v
+\bar{v}K]},
\label{c}
\end{equation}
whose action is given by
\begin{equation}
S_{v} = \int d^{3}x\, \bar{v}\,\hat{O}\, v
\makebox[.5in]{,}
\hat{O} =  \left[- \frac{1}{ 3\kappa} D^2 + \phi(x) + \mu \right].
\end{equation}
Therefore, we obtain,
\begin{eqnarray}
Z_{v,m}&=&\int [\mathcal{D}\phi]\, e^{-S_\phi}\, e^{\int d^3x\, I\left(\frac{\delta~~}{\delta C(x)}\right)} \left. \det \hat{O} 
\int [{\cal D}v][{\cal D}\bar{v}]\, e^{-S_v-\int d^3x\, [\bar{K}v
+\bar{v}K]}\right|_{C=0}\nonumber \\
&=&\int [{\cal D}v][{\cal D}\bar{v}]\, \int [\mathcal{D}\phi]\, e^{-S_\phi} \det \hat{O}\, e^{-S_v-\int d^3x\, \xi\, [v^2+
\bar{v}^2]},
\label{Zvm1}
\end{eqnarray}
Now, in order to obtain the effective theory, we still have to perform the functional integration over
$\phi$. However, the determinant is $\phi$-dependent, so that a closer look to this object is necessary. 
As usual we can write $ \det \hat{O}= e^{\,{\rm tr} \ln \hat{O}}$ and note that ${\rm tr} \ln \hat{O}=F[\phi,\lambda]$ is a functional 
that must be symmetric
under the transformation $\lambda_\mu \rightarrow \lambda_\mu + \partial_\mu \omega$. As there is no parity symmetry breaking, and $\phi$ is real, 
$F$ must depend on $\lambda_\mu$ through the combination $\epsilon_{\mu \nu \rho}\partial_\nu \lambda_\rho$. That is, we can write, 
\begin{equation}
F[\phi,\lambda]=F_\phi[\phi]+F_\lambda[\epsilon \partial \lambda]+F_{\rm int}[\phi,\epsilon \partial \lambda].
\end{equation}
In order to organize a derivative expansion, containing local terms, we can initially suppose $\mu, \kappa \neq 0$. The expansion 
for $ F_\phi[\phi]=\ln \det \left[- \frac{1}{ 3\kappa} \nabla^2 + \mu + \phi(x) \right]$,
will start with the effective potential term, containing no 
derivatives of $\phi$,
\begin{equation}
 F_\phi[\phi]= \int d^3x\, U_{\rm eff}(\phi) +\dots ,
\end{equation}
\begin{eqnarray}
U_{\rm eff}(\phi) &= & \int \frac{d^3k}{(2\pi)^3} \ln \left[(k^2/3\kappa) + \mu +\phi(x) \right]\nonumber \\
&=&  \int \frac{d^3k}{(2\pi)^3} \ln \left[(k^2/3\kappa) + \mu \right] +  \int \frac{d^3k}{(2\pi)^3} \ln \left[1+\frac{\phi}{(k^2/3\kappa) + \mu} \right]\nonumber \\
&=& A+ B\, \phi  -\frac{\phi^2}{2} I_0+\frac{\phi^3}{3} I_1 -\frac{\phi^4}{4} I_2 +\cdots  \nonumber \\
\end{eqnarray}
where $A$ and $B=\int \frac{d^3k}{(2\pi)^3} \frac{1}{(k^2/3\kappa) + \mu}$ are divergent, and $I_n$, $n=0,1,\dots $ are convergent and given by,
\begin{equation}
I_n = \int \frac{d^3k}{(2\pi)^3} \left[ \frac{1}{(k^2/3\kappa) + \mu}\right]^{n+2} = \frac{\kappa^{\frac{3}{2}}}{\mu^{\frac{1}{2}+n}}
\int \frac{d^3u}{(2\pi)^3} \left[\frac{1}{1+u^2/3}\right]^{n+2}.
\end{equation}
The dominant part originated from $F_\lambda[\epsilon \partial \lambda]=\ln \det \left[- \frac{1}{ 3\kappa} D^2 + \mu  \right]$ is a Maxwell term
$\propto \int d^3x\, \frac{1}{2m^2} f_\mu f_\mu $, with $f_\mu=\epsilon_{\mu \nu \rho} \partial_\nu \lambda_\rho$, and $m^2=\kappa \mu$.
After including a linear term in $S_\phi$ and renormalizing, the path integral over $\phi$ can be done by the
replacement,
\begin{equation}
 e^{-S_{\phi} +Tr \ln \hat{O}} \rightarrow  e^{\int d^{3}x\, \left[ B'\, \phi + \frac{\zeta'}{2}\, \phi^2 +\frac{1}{2m^2}\, f^2 \right]}
\end{equation} 
where $\zeta'=\zeta-I_0$, and we have maintained the dominant terms in a large $\mu$ expansion.
Completing the square, now we can perform the integral of the $\phi$ dependent part in eq. (\ref{Zvm1}).
Therefore, the final expression for the partition function of correlated monopoles and center vortices turns out to be,
\begin{eqnarray}
Z_{v,m}&=& {\cal N}\, 
  e^{-\int d^{3}x\, \left\{ \bar{v}\,\left[- \frac{1}{ 3\kappa} D^2 + \mu \right]\, v+ \xi\, [v^2+
\bar{v}^2]+\frac{1}{2\zeta'}\, (\bar{v}{v}-B')^2 +\frac{1}{2m^2}\, f^2 \right\}},
\label{rot}
\end{eqnarray}
The derivation of this partition function is the main result of our work. 
Now, combining eq. (\ref{rot}) with eq. (\ref{W3d}), we obtain the model proposed in refs. \cite{oxman,ldual}, where the nonperturbative sector of 
correlated instantons and center vortices are represented by an effective vortex field, 
\begin{eqnarray}
\lefteqn{Z^{\rm eff}_{YM} = \int [{\cal D}\lambda][\mathcal{D}\Psi][{\cal D}v][{\cal D}\bar{v}]\, e^{-S_{c}}}\nonumber \\
&&\times \,  e^{- \int d^{3}x\,
\left\{ \frac{1}{2}\lambda_{\mu}\lambda_{\mu}-i \lambda_{\mu}k_{\mu}
+ iA_{\mu}
(J_{\mu}^{c}-\epsilon_{\mu\nu\rho}\partial_{\nu}\lambda_{\rho})+ \bar{v}\,\left[- \frac{1}{ 3\kappa} D^2 + \mu \right]\, v+ \xi\, [v^2+
\bar{v}^2]+\frac{1}{2\zeta'}\, (\bar{v}{v}-B')^2 +\frac{1}{2m^2}\, f^2 \right\}}, \nonumber \\
\label{effYM}
\end{eqnarray}
that can be further reduced by keeping the relevant terms when performing the path integral over the $[{\cal D}\Psi]$ sector (see ref. \cite{kondo}),
\begin{eqnarray}
 && Z^{\rm eff} = \int [{\cal D}\lambda][{\cal D}v][{\cal D}\bar{v}]\, e^{- \int d^{3}x\,
\left\{\frac{1}{2} f_\mu \hat{K} f_\mu + \frac{\gamma}{2} \lambda_{\mu}\lambda_{\mu}+ \bar{v}\,\left[- \frac{1}{ 3\kappa} D^2 + \mu \right]\, v
+ \xi\, [v^2+ \bar{v}^2]+\frac{1}{2\zeta'}\, (\bar{v}{v}-B')^2 \right\}}, \nonumber \\
\label{eff}
\end{eqnarray}
where $\hat{K}$ is a differential operator that depends on the Laplacian $\partial^2$, and contains a Maxwell term, 
$\hat{K}=\frac{1}{m^2}+\dots$. 

The vortex sector in eqs. (\ref{effYM}), (\ref{eff}) corresponds to a generalization of the 't Hooft model \cite{hooft1} where an additional 
coupling with the dual field $\lambda_\mu$ has naturally arisen from the calculation. The interesting point regarding this generalization 
is that it allows to relate the different phases of the vortex model with enabled or disabled large dual transformations \cite{ldual}, 
leading to decoupling of the Wilson surfaces or turning them surface variables to be integrated together with the other fields, respectively. 

\section{Conclusions}
\label{conc}

In this article, we have considered three dimensional $SU(2)$
Yang-Mills theory, and followed polymer techniques to derive a field representation of the partition function for the stringlike 
center vortices with monopoles at their borders. For this aim, we have assumed some phenomenological properties such as a vortex stiffness and vortex-vortex interactions. 
In addition, vortices naturally interact with the vector field $\lambda_\mu$ that can be defined in Yang-Mills theories, and that can be thought of as a 
dual field describing the off-diagonal charged sector.

In $SU(2)$, center vortices and monopoles carry magnetic charge $2\pi/g$ and $4\pi/g$, respectively, so that configurations in the ensemble are formed by pairs 
of vortices attached to monopoles and antimonopoles. 
Initially, we have been able to write the ensemble integration in terms of a buiding block $Q(x,x_0)$, the weight 
to be ascribed to the path integral over a center vortex with fixed endpoints and variable length. 
Then, the obtention of the effective theory becomes subject
to the possibility of representing $Q(x,x_0)$ as a vortex field correlator. 
In the noninteracting case, the field representation of the end-to-end
probability for a single stiff polymer is originated from the knowledge of the momenta for this distribution, 
that permits to associate it with a random chain with an effective monomer size. In the interacting case, we had to adopt more
recent techniques developed to study wormlike chains in terms of a Fokker-Plank equation, describing a diffusion $q(x,x_0,u,u_0,L)$ not only in $x$-space 
(the final end-point), but also in $u$-space (the final orientation). After integrating over the lengths, initial and final  
orientations, we obtained an equation for $Q(x,x_0)$, that can be approximated by disregarding components with angular momenta $l\geq 2$
in the $u$-expansion of $q(x,x_0,u,u_0,L)$. In ref. \cite{pol-leshouches}, a similar approximation has been implemented for the 
noninteracting string with stiffness, after associating it with the evolution of a ``rigid body'' in the tangent space. 
This can be justified for semiflexible vortices, as for long chains the probability distribution for the
final orientation is expected to be nearly isotropic. 

As a result of the approximation, the weight $Q(x,x_0)$ turns out to be the Green's function for a Klein-Gordon type operator $\hat{O}$ 
where the usual derivative is replaced by a covariant one, that contains the dual vector field $\lambda_\mu$. 
Finally, by representing this Green's function by means of a complex vortex field, and analyzing 
the dominant terms originated  from the functional determinant $\det \hat{O}$, 
we were able to perform the $\phi$ integration, thus obtaining in a controlled manner a recently proposed effective Abelian model \cite{oxman,ldual} for three 
dimensional $SU(2)$ Yang-Mills theory. In this model, the coupling with the dual vector field is essential to relate the possible phases of the vortex sector 
with enabled or disabled large dual transformations, thus permitting the decoupling, or not, of the Wilson surface appearing in the Petrov-Diakonov 
representation of the Wilson loop \cite{ldual}. This formalism could be extended to accommodate new symmetries such as isospin, and to obtain 
effective field theories for more 
complex systems containing extended objects.

\section{Acknowledgements}

LEO would like to acknowledge C. D. Fosco for fruitful discussions. 
The Conselho Nacional de Desenvolvimento Cient\'{\i}fico e Tecnol\'{o}gico (CNPq-Brazil), PROPPi-UFF and the Funda{\c {c}}{\~{a}}o de Amparo
{\`{a}} Pesquisa do Estado do Rio de Janeiro (FAPERJ) are acknowledged for the
financial support.

\end{document}